# Tunable topological phase transition in soft Rayleigh beam system with imperfect interfaces

Tao Feng [a, †], Letian Gan [a, †], Shiheng Zhao [b], Zheng Chang [b], Siyang Li [a], Yaoting Xue [a], Xuxu Yang [a], Tuck-Whye Wong [a], Tiefeng Li [a, ∗], Weiqiu Chen [a]

[a] Key Laboratory of Soft Machines and Smart Devices of Zhejiang Province and Department of Engineering Mechanics, Zhejiang University, Hangzhou 310027, PR China

[b] College of Science, China Agricultural University, Beijing 100083, PR China

## Abstract

Acoustic metamaterials, particularly the topological insulators, exhibit exceptional wave characteristics that have sparked considerable research interest. The study of imperfect interfaces affect is of significant importance for the modeling of wave propagation behavior in topological insulators. This paper models a soft Rayleigh beam system with imperfect interfaces, and investigates its topological phase transition process tuned by mechanical loadings. The model reveals that the topological phase transition process can be observed by modifying the distance between imperfect interfaces in the system. When a uniaxial stretch is applied, the topological phase transition points for longitudinal waves decrease within a limited frequency range, while they increase within a larger frequency scope for transverse waves. Enhancing the rigidity of the imperfect interfaces also enables shifting of the topological phase transition point within a broader frequency range for longitudinal waves and a confined range for transverse waves. The transition of topologically protected interface modes in the transmission performance of a twenty-cell system is verified, which include altering frequencies, switching from interface mode to edge mode. Overall, this study provides a new approach and guideline for controlling topological phase transition in composite and soft phononic crystal systems.

† These authors contributed equally to this study.
∗ Corresponding author. E-mail address: litiefeng@zju.edu.cn.

# Keywords

Topological phase transition, Topological insulator, Soft Rayleigh beam, Imperfect interface, Interface mode.

# 1. Introduction

Over the past two decades, extensive efforts have been devoted to the study of metamaterials (Ma and Sheng, 2016), which are artificially engineered structures that manifest numerous anomalous phenomena surpassing natural materials. Phononic crystals (PCs), a subcategory of metamaterials, are constituted by periodically arranged constituents or designed utilizing periodic microstructures (Hussein et al., 2014; Jin et al., 2019; Maldovan, 2013). Metamaterials or phononic crystals have catalyzed the development of innovative wave manipulation systems, including negative reflection (Morini et al., 2019; Willis, 2016; Zhang and Liu, 2004), wave attenuation (Barnhart et al., 2019; Chang et al., 2018; Li et al., 2017), cloaking (Chen and Chan, 2007; Chen et al., 2017; Zou et al., 2019), superlens (Aydin et al., 2007; Lin et al., 2009) and energy harvesting devices (Carrara et al., 2013; Chaplain et al., 2020; De Ponti et al., 2021; Hwang and Arrieta, 2022; Lan et al., 2021; Sugino and Erturk, 2018) due to their exceptional and unique wave characteristics.

Particularly, waveguide devices designed based on the conventional metamaterials and PCs are sensitive to structural perturbations such as defects or disorders. Thus, the topological concepts (Yu et al., 2023), such as the quantum Hall effect (QHE) (Chen and Wu, 2016; Haldane, 1988; Thouless et al., 1982; Wang et al., 2015; Zhang et al., 2019), quantum spin Hall effect (QSHE) (Bruene et al., 2012; Kane and Mele, 2005; Miniaci et al., 2018; Weng et al., 2014) and quantum valley Hall effect (QVHE) (Ezawa, 2013; Pal and Ruzzene, 2017; Xiao et al., 2007), have recently been introduced to metamaterials to achieve disorder immunity. A distinguishing feature of topological metamaterials is the existence of topologically protected interface modes (TPIMs) (Prodan and Prodan, 2009; Susstrunk and Huber, 2016) that facilitate robust wave propagation, impervious to backscattering from defects or disorders. This robustness against disorder arises from the bulk-edge correspondence between the topological invariant of the bulk band and the number of edge states. Structures or materials supporting TPIMs are also referred to as topological insulators (TIs) (Ando, 2013; Hasan and Kane, 2010). TPIMs can be achieved in one-dimensional (1D) phononic systems through a combination of PCs with different topological properties in elastic systems (Muhammad et al., 2019; Yin et al., 2018). However, the TPIMs in 1D TIs occur only at singular frequency point (Xiao et al., 2015), leading to a fixed and narrow operating frequency range once designed and fabricated, which constrains their practical applications. Various design of tunable TIs have been proposed to address the limitation, such as employing piezoelectric ceramics (Zhou et al., 2020, 2019) or intelligent magnetoelastic materials (Z. Chen et al., 2022). Those designs, which tune the operating frequency range via external electronic or magnetic fields, are mostly studied in rigid TI systems. When the solid phase of the TIs is soft and highly deformable, more complex wave propagation behaviors can be anticipated under external mechanical loads(Chen et al., 2021; Nguyen et al., 2019). The modeling of wave propagation behavior in deformed TIs is crucial in those designs.

In previous studies that modeling wave propagations and topological phase transition in soft structures (Chen et al., 2021; Y. Chen et al., 2022), the presumption of a continuous state for the displacement and traction vectors across the interface (Jones and Whittier, 1967) has been frequently employed. However, in practical scenarios, the junction between the two distinct solids is often imperfect due to various reasons including damages and defects (Lee and Pyo, 2007; Mahiou and Béakou, 1998). The imperfection may cause relative motion in the tangential and normal directions at the joint, whose impact on wave propagation have been raised in studies of rigid PCs (Li et al., 2021; Yang et al., 2022; Zheng and Wei, 2009). In soft TIs, the achievement of a perfect interface while splicing differently deformed structures is unattainable (Zhou et al., 2020). To enable precise control of the topological phase transition process and provide concrete guidance for designing flexible and practical soft TIs, the study of imperfect interfaces affect is of great significance, but not yet consider.

This paper models a TI system with imperfect interfaces, and studies its topological phase transition process tuned by mechanical loadings. Specifically, the TI system is composed of soft Rayleigh beams with soft thin layer connectors. Initially, the soft thin layers are substituted by a linear spring-layer imperfect interface model as a simplification while studying the propagation of Bloch waves, and its corresponding stiffness under the existence of finite deformation is evaluated. Subsequently, by combining the Raleigh beam theory and imperfect interfaces, the dispersion relation of the beam system is deduced, while a supplementary simplified string model offers insightful estimates and clarifies underlying mechanisms. We further conduct numerical simulations employing the incremental wave motion theory (Ogden, 2007). These simulations assess the accuracy of the imperfect interface simplification and probe the alterations in the band structure and topological properties under differing conditions. Numerical and theoretical findings exhibit good concordance. The periodic tuning of the distance between the imperfect interfaces induces topological phase transition within the system. The transmission ratio of the supercell containing two types of unit cells with different topological properties are also analyzed. A transmission peak occurs in the bandgap frequency range, and the mode shape at this peak confirms the localization of energy, indicating the existence of TPIMs. We demonstrate that the frequency of TPIMs is tunable by applying the uniaxial stretch or modifying the stiffness of the imperfect interfaces. This study can potentially benefit flexible vibration manipulation devices, energy harvesters, and elastodynamic systems.

## 2. Modeling and basic equations

### 2.1. Model description

A schematic of the periodic beam system with its reference configuration $\Omega^c$ in the Cartesian coordinate system is presented in Fig. 1 (a). Each unit cell of the system consists of three beam components: the left sub-beam (1st sub-beam), the middle sub-beam (2nd sub-beam), and the right sub-beam (3rd sub-beam). The 1st and 2nd sub-beam are connected through the thin layer denoted $e_L$ while the 2nd and 3rd sub-beam are connected through $e_R$. Each layer has the initial thickness $d^c$. The length between $e_L$ and $e_R$ is denoted by $2l_t^c$, while the total length of each unit cell is $2l^c$. Note that $d^c \ll l^c$. The height (in the in-plane direction) and width (in the out-of-plane direction) of all beams are $h^c$ and $b^c$, respectively. The material properties of all the beams and the layers are identical, exhibiting a hyperelastic constitution that is homogeneous, isotropic, and soft.

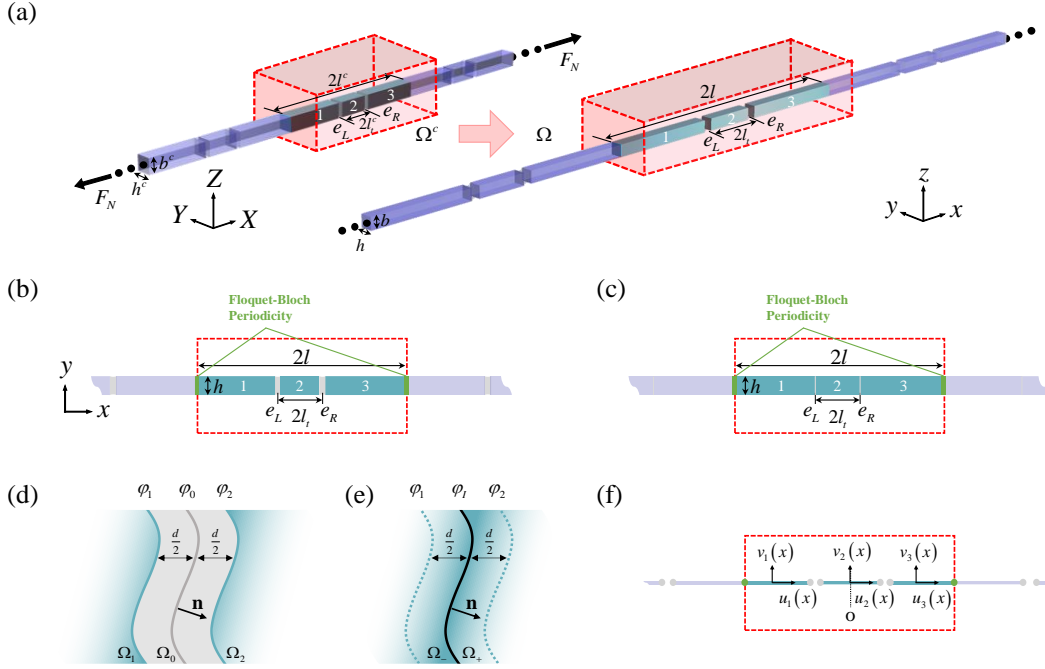

Fig. 1. Schematic diagrams of the soft beam system with imperfect interfaces and its unit cell. (a) The three-dimensional (3D) view of the reference ($\Omega^c$) and current ($\Omega$) configurations before and after the finite deformation. (b) and (c) represent the top view of unit cells in the current configuration adopted for the full and simplified numerical analysis, respectively. (d) The schematic of the three-phase configuration while (e) denotes the two-phase configuration. (f) The schematic diagram for theoretical analysis. The structure of unit cell is boxed in the red rectangle.

The uniaxial stretch $F_N$ is considered as the cause of finite deformation in the beam system, as shown in Fig. 1(a). The finite deformation can be expressed in terms of principal stretches ($\lambda_1$ and $\lambda_2$). As the soft beam system undergoes finite deformation, it expands to the current configuration $\Omega$, where the length of the 2nd sub-beam becomes $2l_t$. Additionally, the total length of each unit cell becomes $2l$, while the height and width become $h$ and $b$, respectively. The thickness of each soft thin layer becomes $d$.

To simplify the analysis of the three-dimensional (3D) problem, we neglect the deformation along the $z$ direction, and adopt a two-dimensional (2D) plane-strain assumption in this paper. The corresponding 2D structure for further in-plane Bloch wave analysis is illustrated in Fig. 1 (b). However, since the width of the beam is significantly shorter than its length, the plane-stress assumption is more appropriate. Consequently, the initial material parameters are substituted to obtain the plane-stress solution from the plane-strain problem (Ugural and Fenster, 2003):

$$\upsilon_{2D} = \frac{2\upsilon_{3D}\mu_{3D}}{\upsilon_{3D} + 2\mu_{3D}}, \ \mu_{2D} = \mu_{3D}, \tag{1}$$

where the parameter $\upsilon$ and $\mu$ represent the first and second *Lamé* constants, respectively. The subscript 3D denotes the material parameter in a 3D (or equivalently 2D plane-stress) problem, while 2D represents the transformed material parameter used in the corresponding 2D plane-strain problem. In the plane-strain assumption, the finite deformation and the incremental wave motion are restricted to the $x-y$ plane. Unless otherwise specified, all material parameters are assumed to be in the 2D condition in the following text.

## 2.2. Finite deformation and incremental wave motion theory

For a hyperelastic solid material with elastic properties characterized by *strain energy function*, denoted by $W = W(\mathbf{F})$, the finite deformation $\chi$ transforms the material particle at point $X_j$ from the reference (undeformed) configuration $\Omega^c$ to point $x_i = \chi(X_i)$ in the current (deformed) configuration $\Omega$. The deformation gradient tensor $\mathbf{F}$ is defined as the gradient of the deformation field, and its component form is given by $F_{ij} = \partial x_i / \partial X_j$. For an unconstrained material, the nominal stress and the Cauchy stress tensors can be expressed as

$$\mathbf{T} = \frac{\partial W}{\partial \mathbf{F}}, \ \boldsymbol{\sigma} = J^{-1}\mathbf{F}\frac{\partial W}{\partial \mathbf{F}}, \tag{2}$$

respectively, where $J = \det \mathbf{F}$ denotes the measurement of volume change. In component forms, Eq. (2) gives

$$T_{ji} = \frac{\partial W}{\partial F_{ij}}, \ \sigma_{ij} = J^{-1}F_{i\alpha}\frac{\partial W}{\partial F_{j\alpha}}. \tag{3}$$

In absence of body force, the equilibrium equation can be written as

$$\nabla_X \cdot \mathbf{T} = 0, \ \nabla_x \cdot \boldsymbol{\sigma} = 0, \tag{4}$$

where $\nabla_X$ and $\nabla_x$ denote the divergence operators relative to $\Omega^c$ and $\Omega$, respectively. On use of Eq. (3), the equation equilibrium Eq. (4) gives

$$C_{piqj}\frac{\partial^2 x_j}{\partial X_p \partial X_q} = 0, \tag{5}$$

where $C_{piqj}$ denotes the components of the elastic tensor $\mathbf{C}$ and are defined through

$$C_{piqj} = \frac{\partial^2 W}{\partial F_{ip} \partial F_{jq}}. \tag{6}$$

Next, we assume a time-dependent infinitesimal incremental wave motion $\boldsymbol{u}(\boldsymbol{x},t) = \dot{\boldsymbol{x}}(\boldsymbol{X},t)$ superimposed on the deformed configuration $\Omega$. The incremental equation of motion becomes

$$\nabla_x \cdot \mathbf{T}_0 = \rho_0 \boldsymbol{u}_{,tt}, \tag{7}$$

where $\mathbf{T}_0 = J^{-1}\mathbf{F}\dot{\mathbf{T}}$ denotes the push-forward counterpart of the incremental stress tensor $\dot{\mathbf{T}} = \mathbf{C}\dot{\mathbf{F}} = \mathbf{C}\nabla_X \dot{\boldsymbol{x}}$ with linear approximation, and $\rho_0 = J^{-1}\rho^c$ represents the current mass density in $\Omega$, with $\rho^c$ denoting the mass density in the reference configuration $\Omega^c$. The subscript 0 indicates the resulting push-forward quantities and the subscript $t$ represents the material time derivative. Then the linearized incremental constitutive law for a compressible hyperelastic material becomes

$$\mathbf{T}_0 = \mathbf{C}_0 \dot{\mathbf{H}}, \tag{8}$$

where $\dot{\mathbf{H}} = \nabla_x \boldsymbol{u} = \nabla_X \dot{\boldsymbol{x}} \mathbf{F}^{-1}$ is the incremental displacement gradient tensor. The fourth-order instantaneous elasticity tensor $\mathbf{C}_0$ is represented in component form by

$$C_{0piqj} = J^{-1} F_{p\alpha} F_{q\beta} C_{\alpha i \beta j}. \tag{9}$$

Then the governing equation of wave motion in the current configuration is expressed as the following component form

$$C_{0piqj} u_{j,pq} = \rho_0 u_{i,tt}, \tag{10}$$

which gives the governing equation of the incremental wave motion in a deformed hyperelastic media.

## 2.3. The stiffness of the soft imperfect interface

Given that the thickness of each soft thin layer is much small comparing to the length of the beam, and the modeling of such small structure complicates numerical computations. For this reason, a substitution model is implemented, as illustrated in Fig. 1 (c). Within this model, the thin layers bonding the sub beams are replaced with the imperfect interfaces. These imperfect interfaces, marked by infinitesimal thickness, link the sub beams and can demonstrate linear elastic, nonlinear, viscoelastic, or viscous behavior (Abrate and Di Sciuva, 2018). Of all the models available to portray imperfect interfaces, the linear spring-layer interface model is arguably the simplest and most wildly used. Motivated by this observation, we adopt a linear spring-layer interface model in our Bloch wave analysis, functioning as a linear spring and disrupting displacement continuity in both tangential and normal directions.

The stiffness of the imperfect interface can be derived through the equivalent change from the three-phase configuration (Fig. 1 (d)) to the two-phase configuration (Fig. 1 (e)). The three-phase configuration represents the connecting situation of Fig. 1 (d), where the sub beams, denoted as region $\Omega_1$ and $\Omega_2$, are connected by a thin and soft layer denoted as region $\Omega_0$. The interfaces, denoted as $\varphi_1$ and $\varphi_2$ between the sub beams and the layer are perfect, which indicates that the displacement and the traction are continuous. As inspired by previous research (Benveniste, 2006; Zhu et al., 2011), the displacement $u_i^{(0)}$ evaluated on $\varphi_0$ admit the following Taylor expansions:

$$u_i^{(0)}\Big|_{\varphi_0} = u_i^{(0)}\Big|_{\varphi_1} + \frac{d}{2}u_{i,j}^{(0)}n_j\Big|_{\varphi_1} + o(d^2),$$
$$u_i^{(0)}\Big|_{\varphi_0} = u_i^{(0)}\Big|_{\varphi_2} - \frac{d}{2}u_{i,j}^{(0)}n_j\Big|_{\varphi_2} + o(d^2),$$
(11)

where the notation $u_i^{(p)}\Big|_{\varphi_q}$ represents the displacement in region $\Omega_p$ and on boundary $\varphi_q$, $n_i$ represents the unit normal vector field and $o(d^2)$ representing the terms of orders higher than $d$. From these expressions it follows that

$$u_i^{(2)}\Big|_{\varphi_2} - u_i^{(1)}\Big|_{\varphi_1} = u_i^{(0)}\Big|_{\varphi_2} - u_i^{(0)}\Big|_{\varphi_1} = \frac{d_e}{2}\left(u_{i,j}^{(0)}n_j\Big|_{\varphi_1} + u_{i,j}^{(0)}n_j\Big|_{\varphi_2}\right) + o(d^2). \quad (12)$$

Next, the three-phase configuration is substituted by the two-phase configuration (Fig. 1 (e)), where the layer $\Omega_0$ is eliminated, $\Omega_1$ and $\Omega_2$ are extended to the middle surface $\varphi_0$. Thus, the interface $\varphi_I$ between the extended $\varphi_1$ and $\varphi_2$ coincides geometrically with $\varphi_0$ but is mechanically imperfect. To find the jump of the displacement field across the imperfect interface, we apply Taylor's expansion to the two-phase configuration again and then get

$$u_i^{(1)}\Big|_{\varphi_1} = u_i^{(-)}\Big|_{\varphi_I} - \frac{d_e}{2}u_{i,j}^{(-)}n_j\Big|_{\varphi_I} + o(d_e^2),$$
$$u_i^{(2)}\Big|_{\varphi_2} = u_i^{(+)}\Big|_{\varphi_I} + \frac{d_e}{2}u_{i,j}^{(+)}n_j\Big|_{\varphi_I} + o(d_e^2),$$
(13)

where the notation $\Omega_-$ and $\Omega_+$ represent the extended region of $\Omega_1$ and $\Omega_2$. Then we get

$$u_i^{(2)}\Big|_{\varphi_2} - u_i^{(1)}\Big|_{\varphi_1} = u_i^{(+)}\Big|_{\varphi_I} - u_i^{(-)}\Big|_{\varphi_I} + \frac{d_e}{2}\left(u_{i,j}^{(+)}n_j\Big|_{\varphi_I} + u_{i,j}^{(-)}n_j\Big|_{\varphi_I}\right). \quad (14)$$

Combining Eq. (12) and Eq. (14), we get

$$\llbracket u_i \rrbracket = u_i^{(+)}\Big|_{\varphi_I} - u_i^{(-)}\Big|_{\varphi_I} = \frac{d_e}{2}\left(u_{i,j}^{(0)}n_j\Big|_{\varphi_1} + u_{i,j}^{(0)}n_j\Big|_{\varphi_2} - u_{i,j}^{(+)}n_j\Big|_{\varphi_I} - u_{i,j}^{(-)}n_j\Big|_{\varphi_I}\right), \quad (15)$$

where the symbol $\llbracket u_i \rrbracket$ denotes the jump of the displacement field. The terms $u_{i,j}^{(0)}n_j\Big|_{\varphi_1}$ and $u_{i,j}^{(0)}n_j\Big|_{\varphi_2}$ are to be expressed by quantities in the two-phase configuration.

First, decompose $\nabla \boldsymbol{u}$ into two orthogonal directions (Gu, 2008). Defining $\mathbf{V}_n = \boldsymbol{n} \otimes \boldsymbol{n}$, $\mathbf{V}_s = \mathbf{I} - \boldsymbol{n} \otimes \boldsymbol{n}$, then

$$\nabla \boldsymbol{u} = \nabla_n \boldsymbol{u} + \nabla_s \boldsymbol{u} = \nabla \boldsymbol{u} \mathbf{V}_n + \nabla \boldsymbol{u} \mathbf{V}_s, \quad (16)$$

the component form of Eq. (8) yields

$$T_i = C_{0jilk}n_j u_{k,l} = C_{0jilk}n_j\left(\nabla \boldsymbol{u} \mathbf{V}_n + \nabla \boldsymbol{u} \mathbf{V}_s\right)_{kl} = C_{0jilk}n_j n_l u_{k,t} n_t + C_{0jilk}n_j\left(\nabla_s \boldsymbol{u}\right)_{kl}, \quad (17)$$

defining $Q_{ij} = C_{0piqj}n_p n_q$, which also denotes the *Christoffel acoustic tensor*, and $\mathbf{P} = \mathbf{Q}^{-1}$, $\Lambda_{ijk} = P_{im}n_q C_{0qmkj}$, then Eq. (17) becomes

$$u_{i,j}n_j = P_{ij}T_j - \Lambda_{ijk}\left(\nabla_s \boldsymbol{u}\right)_{jk}, \quad (18)$$

and the terms in Eq. (15) can be expressed as

$$\begin{aligned}
u_{i,j}^{(0)}n_j\Big|_{\varphi_1} &= P_{ij}^{(0)}T_j^{(1)}\Big|_{\varphi_1} - \Lambda_{ijk}^{(0)}\left(\nabla_s \boldsymbol{u}^{(1)}\right)_{jk}\Big|_{\varphi_1}, \\
u_{i,j}^{(0)}n_j\Big|_{\varphi_2} &= P_{ij}^{(0)}T_j^{(2)}\Big|_{\varphi_2} - \Lambda_{ijk}^{(0)}\left(\nabla_s \boldsymbol{u}^{(2)}\right)_{jk}\Big|_{\varphi_2}, \\
u_{i,j}^{(+)}n_j\Big|_{\varphi_I} &= P_{ij}^{(2)}T_j^{(+)}\Big|_{\varphi_I} - \Lambda_{ijk}^{(2)}\left(\nabla_s \boldsymbol{u}^{(+)}\right)_{jk}\Big|_{\varphi_I}, \\
u_{i,j}^{(-)}n_j\Big|_{\varphi_I} &= P_{ij}^{(1)}T_j^{(-)}\Big|_{\varphi_I} - \Lambda_{ijk}^{(1)}\left(\nabla_s \boldsymbol{u}^{(-)}\right)_{jk}\Big|_{\varphi_I}.
\end{aligned} \quad (19)$$

For simplicity, we dismiss the mass of the thin layers and assume that the traction is continuous: $T_j^{(1)} = T_j^{(2)} = T_j^{(+)} = T_j^{(-)} = T_j$. By substituting Eq. (19) into Eq. (15), we get:

$$\llbracket u_i \rrbracket = \frac{d}{2}\left[\left(2P_{ij}^{(0)} - P_{ij}^{(1)} - P_{ij}^{(2)}\right)T_j - \left(\Lambda_{ijk}^{(0)} - \Lambda_{ijk}^{(2)}\right)\left(\nabla_s \boldsymbol{u}^{(+)}\right)_{jk} - \left(\Lambda_{ijk}^{(0)} - \Lambda_{ijk}^{(1)}\right)\left(\nabla_s \boldsymbol{u}^{(-)}\right)_{jk}\right]. \quad (20)$$

In this paper, we consider the case that the material of the thin layers is much softer

than the surrounding material, then we get

$$d = \varepsilon \hat{d}, \ C_0^{(0)} = \varepsilon \hat{C}_0^{(0)}, \ C_0^{(1)} = C_0^{(2)} = \hat{C}_0^{(1)} = \hat{C}_0^{(2)}, \tag{21}$$

where $\varepsilon$ is a small positive dimensionless parameter, $\hat{\cdot}$ denotes the typical quantities with $\hat{C}_0^{(0)} \approx \hat{C}_0^{(1)} = \hat{C}_0^{(2)}$, then

$$P_{ij}^{(0)} = \frac{1}{\varepsilon} \hat{P}_{ij}^{(0)}, \ \Lambda_{ijk}^{(0)} = \hat{\Lambda}_{ijk}^{(0)}, \ P_{ij}^{(1)} = \hat{P}_{ij}^{(1)},$$
$$P_{ij}^{(2)} = \hat{P}_{ij}^{(2)}, \ \Lambda_{ijk}^{(1)} = \hat{\Lambda}_{ijk}^{(1)}, \ \Lambda_{ijk}^{(2)} = \hat{\Lambda}_{ijk}^{(2)}. \tag{22}$$

Substituting Eq. (21) and Eq. (22) into Eq. (20), we get

$$\llbracket u_i \rrbracket = d P_{ij}^{(0)} T_j + o(\varepsilon), \tag{23}$$

where $o(\varepsilon)$ representing the terms of orders higher than $\varepsilon$. Then the jump condition of the imperfect interface becomes

$$T_j = \left(d P_{ij}^{(0)}\right)^{-1} \llbracket u_i \rrbracket = \frac{Q_{ij}^{(0)}}{d} \llbracket u_i \rrbracket = K_{ij} \llbracket u_i \rrbracket, \tag{24}$$

then the expression of the stiffness of the soft imperfect interface yields

$$K_{ij} = \frac{1}{d} Q_{ij}^{(0)} = \frac{1}{d} C_{0piqj}^{(0)} n_p n_q. \tag{25}$$

## 2.4. In-plane Bloch wave analysis

To demonstrate the influence of imperfect interfaces and finite deformation on the band structure, a theoretical model for the Bloch wave propagating in a unit cell comprising three Rayleigh beams and two imperfect interfaces is established, as the schematic diagram shown in Fig. 1 (f). Each Rayleigh beam contains two types of in-plane elastic waves, namely the longitudinal (P-) wave and the transverse (S-) wave, their time-harmonic governing equations are given by (Zhao and Chang, 2021)

$$E_0 u_i''(x) + \rho_0 \omega^2 u_i(x) = 0,$$
$$P_0 E_0 v_i''''(x) + \left(P_0 \rho_0 \omega^2 - G_0\right) v_i''(x) - \rho_0 \omega^2 v_i(x) = 0, \tag{26}$$

where $u$ and $v$ denote the displacement along the $x$ or $y$ direction, respectively, and the subscript $i$ denotes the number of the Rayleigh beam, as shown in Fig. 1 (c). $E_0$ represents the effective Young's modulus, $G_0$ represents the effective shear modulus and $P_0$ represents the moment of inertia per area. These parameters are defined as (Zhao and Chang, 2021)

$$E_0 = C_{01111} - \frac{C_{01122}C_{02211}}{C_{02222}}, \quad G_0 = C_{01212} - \frac{C_{01221}C_{02112}}{C_{02121}}, \quad P_0 = \frac{I_0}{A_0} = \frac{1}{12}h^2. \tag{27}$$

The internal normal force $N_i$, bending moment $M_i$ and shear force $V_i$ can be expressed as

$$\begin{aligned} N_i(x) &= E_0 A_0 u_i'(x), \quad M_i(x) = E_0 I_0 v_i''(x), \\ V_i(x) &= -E_0 I_0 v_i'''(x) + \left(G_0 A_0 - \rho_0 I_0 \omega^2\right) v_i'(x). \end{aligned} \tag{28}$$

The general solution of Eq.(26) is

$$\begin{aligned} u_i(x) &= A_i^1 \cos(\zeta x) + A_i^2 \sin(\zeta x), \\ v_i(x) &= B_i^1 \cosh(\alpha x) + B_i^2 \sinh(\alpha x) + B_i^3 \cos(\beta x) + i B_i^4 \sin(\beta x), \end{aligned} \tag{29}$$

where $i$ denotes $\sqrt{-1}$, $A_i^j$, $B_i^j$ denotes the $j$-th real constant on the $i$-th beam, and parameters $\alpha$, $\beta$, $\zeta$ are defined as:

$$\begin{aligned} \alpha &= \sqrt{\frac{G_0 - P_0 \rho_0 \omega^2 + \gamma}{2P_0 E_0}}, \quad \beta = \sqrt{\frac{-G_0 + P_0 \rho_0 \omega^2 + \gamma}{2P_0 E_0}}, \\ \gamma &= \sqrt{\left(G_0 - P_0 \rho_0 \omega^2\right)^2 + 4P_0 E_0 \rho_0 \omega^2}, \quad \zeta = \sqrt{\frac{E_0}{\rho_0}}\omega. \end{aligned} \tag{30}$$

For a unit cell containing 3 Rayleigh beams, there exist 18 undetermined constants $A_1^1 - A_3^2$ and $B_1^1 - B_3^4$. In order to solve such system, 18 boundary conditions are necessary, which can be found through 6 Floquet-Bloch periodic conditions, 6 stress continuity conditions and 6 displacement discontinuity conditions induced by the imperfect interface, which includes the following:

1) 6 Floquet-Bloch periodic conditions, i.e.,

$$\Psi(x + R) = \Psi(x) e^{i k \cdot R}, \tag{31}$$

representing the translational periodicity of the space function $\Psi$ in a system with spatial periodicity. For particular, in present work, Eq. (31) yields

$$\begin{aligned} u_3(l) &= u_1(-l)e^{2ikl}, \quad v_3(l) = v_1(-l)e^{2ikl}, \quad v_3'(l) = v_1'(-l)e^{2ikl}, \\ N_3(l) &= N_1(-l)e^{2ikl}, \quad M_3(l) = M_1(-l)e^{2ikl}, \quad V_3(l) = V_1(-l)e^{2ikl}, \end{aligned} \tag{32}$$

representing the periodicity of longitudinal and transverse displacement, angle of rotation, internal normal force, bending moment and shear force, respectively. $i$ denotes $\sqrt{-1}$, $k$ denotes the Bloch wave vector, for a 1D problem with unit cell length $2l$, it varies in the first Brillouin zone $[-\pi/2l, \ \pi/2l]$.

2) 6 stress continuity conditions, i.e.,

$$N_1(-l_t) = N_2(-l_t), \ N_2(l_t) = N_3(l_t),$$
$$M_1(-l_t) = M_2(-l_t), \ M_2(l_t) = M_3(l_t), \quad (33)$$
$$V_1(-l_t) = V_2(-l_t), \ V_2(l_t) = V_3(l_t),$$

representing the continuity of internal normal force, bending moment and shear force, respectively. Such condition is established based on the hypothesis that the mass of the imperfect interface is negligible, which is a reasonable assumption considering that the thicknesses of the soft thin layers are neglected.

3) 6 displacement discontinuity conditions induced by the imperfect interface, i.e.,

$$N_2(-l_t) = k_N [u_2(-l_t) - u_1(-l_t)], \ N_3(l_t) = k_N [u_3(l_t) - u_2(l_t)],$$
$$M_2(-l_t) = k_M [v_2'(-l_t) - v_1'(-l_t)], \ M_3(l_t) = k_M [v_3'(l_t) - v_2'(l_t)], \quad (34)$$
$$V_2(-l_t) = k_V [v_2(-l_t) - v_1(-l_t)], \ V_3(l_t) = k_V [v_3(l_t) - v_2(l_t)],$$

representing the relationship between the internal force and displacement discontinuity. Define

$$\tilde{k}_r = \frac{k_r}{A_0}, \ (r = M, \ N, \ V) \quad (35)$$

as the spring constant per unit area of the imperfect interface, which can be defined through the comparison of Eq. (24) and Eq. (34).

Eqs. (32)-(34) provides the complete set of boundary conditions that govern the in-plane Bloch wave in the Rayleigh beam system with imperfect interfaces. Substituting Eq. (29) into Eqs. (32) - (34) leads to an algebraic homogeneous linear system. Decoupling the constants $A_1^1 - A_3^2$, $B_1^1 - B_3^4$ from each other, it can be split into 2 groups, i.e.,

$$p(\omega, k)a = 0,$$
$$q(\omega, k)b = 0, \quad (36)$$

where $p$ is a $6 \times 6$ and $q$ is a $12 \times 12$ coefficient matrix. Vector $a$ and $b$ collects all of the real constants $A_1^1 - A_3^2$, $B_1^1 - B_3^4$, respectively. The non-trivial solution of Eq. (36) can be found in condition that

$$\det p(\omega, k) = 0,$$
$$\det q(\omega, k) = 0. \quad (37)$$

Hence, the $\omega - k$ relationship, also called the *dispersion relation*, can be deduced from Eq. (37). For each given $k$ or angular frequency $\omega$, vector $a$ and $b$ can be calculated from Eq. (36). Eq. (36) and (37) determine the in-plane Bloch wave propagation in Rayleigh beam with imperfect interfaces.

A large initial slenderness ratio might lead to $P_0 \to 0$, the governing equation of the transverse wave degrade from the Rayleigh beam model Eq. (26)$_2$ into the string model

$$G_0 v_i''(x) + \rho_0 \omega^2 v_i(x) = 0, \tag{38}$$

which has a similar form to the longitudinal wave. The shear force Eq. (28)$_3$ becomes

$$V_i(x) = G_0 A_0 v_i'(x). \tag{39}$$

Following the same procedure as Eq. (29)-(36), and let $\tilde{k}_p = \tilde{k}_N, \tilde{k}_s = \tilde{k}_V$, then the expression of the *dispersion relation* Eq. (37) for the string model yields

$$\kappa^2 R_m(\omega, k) + 2\kappa \sin(\eta_m \omega) + T_m(\omega) = 0, \tag{40}$$

with $m = p$ for the longitudinal (P-) and $m = s$ for the transverse (S-) wave mode, in which

$$R_m(\omega, k) = \cos(2lk) - \cos(\eta_m \omega), \quad T_m(\omega) = \cos(\eta_m \omega) - \cos\left(\eta_m \omega - \frac{4l_t \omega}{c_m}\right),$$

$$\eta_m = \frac{2l}{c_m}, \quad c_p = \sqrt{\frac{E_0}{\rho_0}}, \quad c_s = \sqrt{\frac{G_0}{\rho_0}}, \quad \kappa = \frac{2\tilde{k}_m}{c_m \rho_0 \omega}. \tag{41}$$

For further simplification, a large stretch might lead to $G_0 \to E_0$ as reported in the existing study on a soft lattice (Zhao et al., 2022), which leads to $c_s \to c_p$, and if $\tilde{k}_p = \tilde{k}_s$, then Eq. (40) become identical for either $m = p$ or $s$, which indicates an overlap of the band curves under the longitudinal and the transverse wave mode. Detailed information about such simplification can be referred to the Supplementary Material.

## 2.5. *Calculation of Zak phase*

Topological properties of TIs cannot be comprehensively ascertained solely through the *dispersion relation*. Instead, topological invariants are employed to characterize these attributes. In 1D systems, the Zak phase is typically utilized to convey information regarding the edge-mode state of specific bands. The Zak phase was originally developed for electronic systems as a special type of Berry phase (Zak,

1989), which is an essential geometric phase factor that plays an important role in many physical phenomena (Atala et al., 2013; Delplace et al., 2011; Kameda et al., 2019). Latterly, it has been extended to photonic (Gao et al., 2015; Longhi, 2013), acoustic (Xiao et al., 2015) and elastic systems (Cajic et al., 2021; Chaunsali et al., 2017; Muhammad et al., 2019). The significance of the Zak phase is self-explanatory, including identification of the topological properties and existence of TPIMs.

For a unit cell exhibiting mirror symmetry relative to its central cross-sectional plane, the Zak phase is constrained to only two possible values: 0 or $\pi$. This distinction is critical in identifying the topological properties and determining the existence of topologically protected interface modes. If the Zak phase equals 0, it indicates that the edge-mode shape at both the center ($k=0$) and edge ($k=\pm\pi/2l$) of the first Brillouin zone possesses identical symmetry, which can be either symmetric or antisymmetric. Conversely, if the Zak phase equals $\pi$, it indicates that the edge-mode shape at the center and edge of the first Brillouin zone possesses opposite symmetry.

In current study, the Zak phase is calculated through numerical methods. The Zak phase is defined as an integration of the Berry connection over the first Brillouin zone, which can be mathematically expressed as (Xiao et al., 2015)

$$\theta_n^{\text{zak}} = i \oint dk \int_\Omega \left[ w_n^*(k) \partial_k w_n(k) \right] dr, \quad (42)$$

where the subscript *n* denotes the *n*-th band curve and *w* denotes the periodic-in-cell mode function, i.e., $u_n$ or $v_n = e^{ikx} w_n$, with *u* and *v* denotes the normal mode function of longitudinal and transverse wave, respectively. The superscript asterisk denotes conjugation, and $\Omega$ denotes the region of unit cell. To facilitate computation, the discrete form of Eq. (42) yields (Cajic et al., 2021)

$$\theta_n^{\text{zak}} = -\text{Im}\left\{ \sum_{i=-N}^{N-1} \ln\left[ w_n^*\left(\frac{i}{N}\pi\right) w_n\left(\frac{i+1}{N}\pi\right) \right] \right\}, \quad (43)$$

where Im denotes the imaginary part and *N* denotes the total number of the discrete Bloch wave vector *k*.

After obtaining the Zak phase for each band curve, the topological property of the *p*-th bandgap can be determined through the bandgap sign parameter $\varsigma$ (Xiao et al., 2014), which is defined as

$$\text{sgn}(\varsigma_p) = (-1)^p (-1)^t \exp\left( i \sum_{n=1}^{p} \theta_n^{\text{zak}} \right), \quad (44)$$

where *t* is an integer indicates that the number of band crossing points beneath the *p*-th bandgap. As will be explained later, the existence of TPIMs can be determined through the use of Eq. (44).

## 2.6. FE numerical analysis

To investigate the effect of finite deformation on the change of the stiffness of the soft imperfect interface and the incremental wave motion, two types of the numerical models are established through the finite-element-method-based software COMSOL Multiphysics 6.1.

The first type is referred to as the full numerical model, which includes the physical modeling of imperfect interfaces and needs several steps. At first step, a nonlinear quasi-static analysis with a 2D plane-strain assumption is carried out in the solid mechanics module to calculate the finite deformation of a unit cell conforming to Eq. (4). Subsequently, the deformation gradient acquired in the initial step are used to determine the instantaneous elasticity tensor $\mathbf{C}_0$ through Eq. (9), and the deformed geometry (Fig. 1(b)) is also used for further eigenvalue analysis.

The second type of the numerical model is referred to as the simplified numerical model. At first step, the deformation gradient of each part is calculated theoretically with various assumptions applied for simplicity. Next, the numerical model in accord with schematic Fig. 1(c) is built. The geometric and material parameters is determined through the deformation gradient obtained at the first step. Moreover, to introduce elastic discontinuity conditions between adjacent beams, the Thin Elastic Layer node is employed on the boundary pair, accommodating the assumption of imperfect interfaces. The stiffness of the Thin Elastic Layer is calculated through Eq. (25).

The final step of both types of the numerical models is applying the Floquet-Bloch periodic condition Eq. (31) on the left and right ends of the unit cell, as depicted in Fig. 1 (b) and (c). Then the Bloch wave vector *k* is swept over the first Brillouin zone $[-\pi/2l,\ \pi/2l]$, and for each *k*, the corresponding $\omega$ is found via eigenvalue analysis of Eq. (10) in the weak-form PDE module to obtain the $k-\omega$ relationship, i.e., the *dispersion relation*. Comprehensive details on these numerical setups and meshes are presented in the Supplementary Material. It's clear from this information that the simplified numerical model has superior mesh quality, needs fewer meshes and degrees of freedom (DOFs), and is thus more efficient.

To differentiate the wave mode of each band branch from the numerical outcomes, the amount of polarization (Achaoui et al., 2010) is utilized:

$$p^2 = \frac{\int_\Omega |u|^2 \, dr}{\int_\Omega (|u|^2 + |v|^2) \, dr}, \tag{45}$$

the integral takes over the current configuration of the unit cell. For the in-plane wave motion, $p^2 = 0$ or $1$ represents the longitudinal or transverse wave mode, respectively (abbreviated as Mode P or S). In this study, we have employed the range of $0.8 \leq p^2 \leq 1$ for the red color and $0 \leq p^2 \leq 0.2$ for the blue color to distinguish between these two types of wave motion. Notably, there exists no coupled mode that results in a value of $p^2$ between 0.2 and 0.8.

## 3. Results and discussions

In this section, both theoretical and numerical results are presented to evaluate the tunability of the topological phase transition process through imperfect interfaces, as well as the influence of finite deformation on the dispersion relation and topological characteristics of the soft Rayleigh beam system. Specifically, we employ the Neo-Hookean hyperelastic model as the material model of the Rayleigh beams and the thin layers, with the corresponding *strain energy function* defined as

$$W(\mathbf{F}) = \frac{\upsilon}{2}(J-1)^2 - \mu \ln(J) + \frac{\mu}{2}\left[\text{tr}(\mathbf{F}^T \mathbf{F}) - 2\right]. \tag{46}$$

In the following calculations, set the geometric parameters of the undeformed unit cell as: $l^c = 20 h^c = 0.5 \text{ m}$, $l_t^c = \delta l^c$ where $\delta$ is a structural parameter ranging from 0.02 to 0.44. The thickness of each thin layer is $d^c = 0.3 \text{ mm}$. The initial material parameters, including the *Lamé* constants of the soft Rayleigh beam are chosen as $\upsilon^m = 1.5 \text{ MPa}$, $\mu^m = 1 \text{ MPa}$, while for the thin layers $\upsilon^e = 1 \text{ kPa}$, $\mu^e = 9 \text{ kPa}$. The superscript *m* and *e* on *Lamé* constants denote the Rayleigh beam and the thin layer, respectively. For such chosen strain energy function, the explicit expression of Eq. (25) yields

$$\tilde{k}_N = \frac{C_{01111}^{(0)}}{d} = \frac{\upsilon^e (\lambda_1^e \lambda_2^e)^2 + \mu^e (1 + (\lambda_1^e)^2)}{d^c (\lambda_1^e)^2 \lambda_2^e}, \quad \tilde{k}_V = \frac{C_{01212}^{(0)}}{d} = \frac{\mu^e}{d^c \lambda_2^e}, \quad \tilde{k}_M = P_0 \tilde{k}_N, \tag{47}$$

under uniaxial stretch, where the superscript $e$ on $\lambda_i$ denotes the principle stretches of each thin layer. Moreover, under the free deformation condition, i.e., $\lambda_1^e = \lambda_2^e = 1$, the stiffness of the imperfect interface Eq. (47) yields

$$\tilde{k}_N = \frac{\upsilon^e + 2\mu^e}{d^c} = 6.33 \times 10^7 \text{ N/m}^3, \quad \tilde{k}_V = \frac{\mu^e}{d^c} = 3 \times 10^7 \text{ N/m}^3, \tag{48}$$

which is in accord with previous research (Mal et al., 1989; Pilarski and Rose, 1988). In addition, the density for both the Rayleigh beams and the thin layers is $\rho^c = 1110 \text{ kg/m}^3$. For simplicity, we neglect the material damping.

### 3.1. The comparison of different models under finite deformation

The comparison between the simplified numerical and theoretical model with the full numerical model under different uniaxial stretch is presented in Fig. 2. Without loss of generality, we set $\delta = 0.1$. The solid dot lines denote the full numerical (FE) model, dashed dot lines denote the simplified numerical model, while solid lines denote theoretical results.

In the deformation free configuration ($\lambda_1 = 1$), outstanding congruence between the results of theoretical and numerical analysis is evident from Fig. 3. As the uniaxial stretch increases, a slight discrepancy between the simplified numerical model and the full numerical model appears. However, this difference remains minor and within our acceptable tolerance levels. Simultaneously, the disparity between the simplified numerical model and the theoretical model becomes even less significant.

The simplified numerical model ensures the correctness about the equivalence of imperfect interfaces, especially in the condition that the pre-deformation is small, and the theoretical model effectively captures the fundamental characteristics of Bloch wave propagation within the beam system. For ease of calculation, in following sections, we mainly employ the simplified model for numerical calculations.

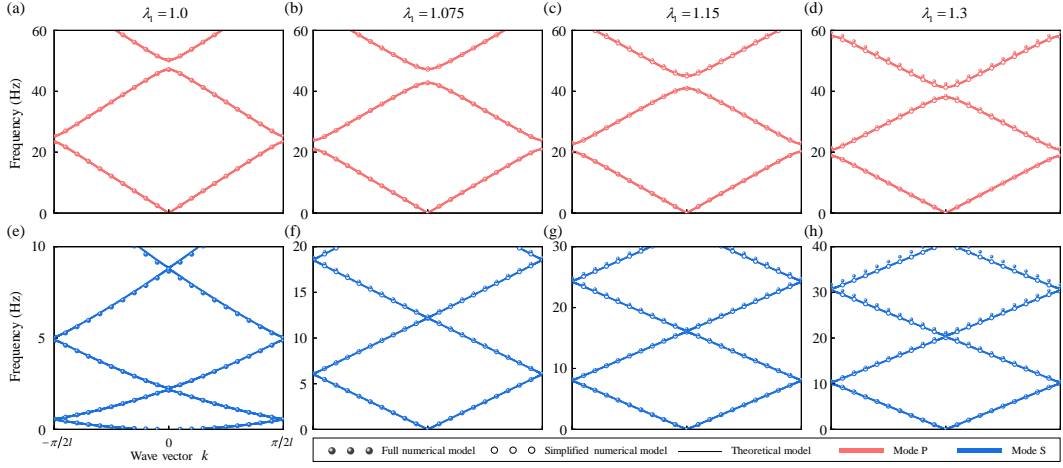

Fig. 2. The comparison between the full (solid dot lines), simplified (dashed dot lines) numerical model and the theoretical model (solid lines). (a)-(d) The longitudinal wave. (e)-(h) The transverse wave.

## *3.2. Topological phase transition*

The band structure of the Rayleigh beam system with imperfect interfaces is illustrated in Fig. 3 under the following conditions: $\lambda_1 = \lambda_2 = 1$, $\tilde{k}_N = 6.33 \times 10^7 \text{ N/m}^3$, $\tilde{k}_V = 3 \times 10^7 \text{ N/m}^3$ and $\tilde{k}_M = P_0 \tilde{k}_N$ with different geometrical parameter $\delta$ in the first Brillouin zone. Fig. 3 (a) - (c) depict the band inversion process for the longitudinal wave, whereas Fig. 3 (d) - (f) display that for the transverse wave. Mode P presents the first three band branches from 0 to 60 Hz, while Mode S offers the first four band branches from 0 to 10 Hz. The discrepancy in the studied frequency range of band structure arises from the fact that longitudinal waves exist at higher frequencies compared to transverse waves. Consequently, eigenmodes at elevated frequency ranges should be calculated for longitudinal wave. In this study, we examine the soft beam system with varying $\delta$ values representing distinct configurations with consistent initial length.

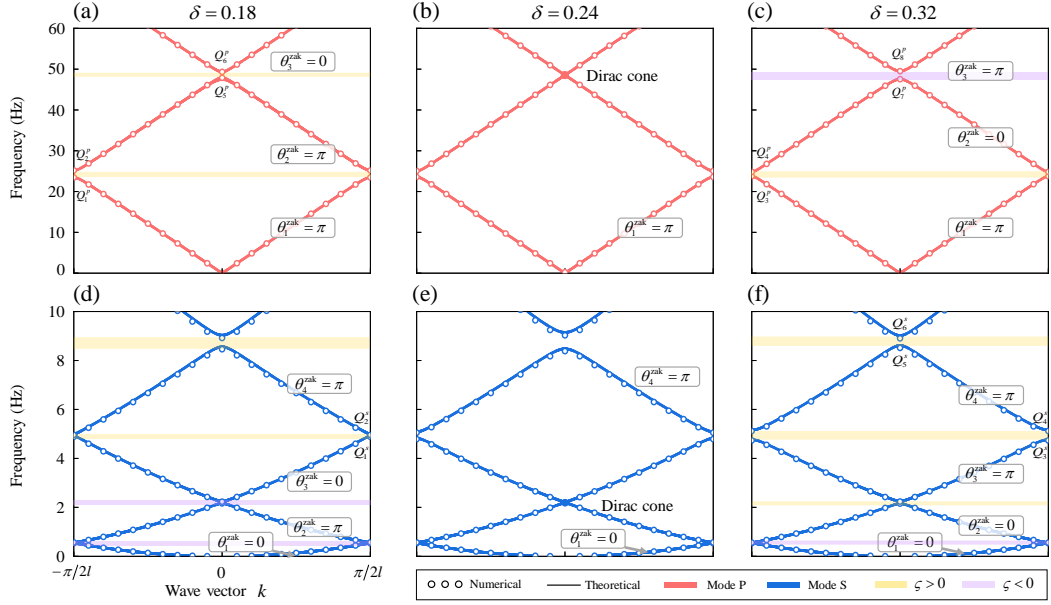

Fig. 3. Band structures of in-plane wave for the soft Rayleigh beam system with imperfect interfaces under different geometrical parameter $\delta$ in the first Brillouin zone. (a) - (c) Topological transition process of the longitudinal wave (red lines), while (d) - (f) the same process of the transverse wave (blue lines) for three different values of $\delta = 0.18$, 0.24, 0.32, respectively; The dashed dot and solid lines denote the simplified numerical and theoretical results, respectively. The Zak phases of each band branch is labeled by grey textbox. The stripes in yellow and purple indicate the bandgap signs with $\varsigma > 0$ and $\varsigma < 0$, respectively.

From Fig. 3 (a) - (c), the second bandgap of Mode P closes at the center of the Brillouin zone and a topological transition point marked by a linear crossover, denoted as the Dirac cone, occurs with the increase of the geometrical parameter $\delta$ from 0.18 to 0.24. This closure of the bandgap occurs when the sub-beams are not equally divided in the unit cell, which is a notable difference from the corresponding result for longitudinal waves (Chen et al., 2021). The second bandgap reopens when the geometrical parameter $\delta$ increases from 0.24 to 0.32. Fig. 3 (d) - (f) depicts a similar band inversion process for Mode S, and the Dirac cone forms when $\delta = 0.24$, which is the same as the state observed in Mode P.

The Zak phase for each branch of the bands is also calculated through Eq. (43) and represented in Fig. 3 (a) - (f). During the closing and reopening process of the second bandgap, the Zak phase of the second and third branches, which forms the second bandgap, become inversed with respect to each other regardless of the polarization mode (P or S). Specifically, the Zak phase of the second band curve changes from $\pi$ at $\delta = 0.18$ to 0 at $\delta = 0.32$ after the formation of the Dirac cone, while the Zak phase of the third band curve transforms from 0 at $\delta = 0.18$ to $\pi$ at $\delta = 0.32$. However, for Mode P, the Zak phase of the first band branch remains $\pi$, whereas for Mode S, it remains zero regardless of changes in $\delta$. In summary, the Zak phases of the second and third band branches change after the band crossing, providing evidence for the existence of a topological phase transition process.

After obtaining the Zak phase of each band branch, the bandgap sign parameter $\varsigma$ can be determined using Eq. (44) and is represented in Fig. 3, providing verification for the existence of TPIMs. The yellow and purple regions in Fig. 3 denote bandgap signs with $\varsigma > 0$ and $\varsigma < 0$, respectively, to demonstrate different topological properties of the bandgaps. It can be observed from Fig. 3 (a)-(f) that the sign of the second bandgap changes after the topological phase transition, while the signs of the other bandgaps remain the same. This indicates that there exists an interface mode within the range of the second bandgap.

### *3.3. Topologically protected interface mode*

We analyze the topological phase transition process for varying geometrical parameter $\delta$ from 0.02 to 0.44. The *topological phase diagram* is shown in Fig. 4. Fig. 4 (a) illustrates the evolution of the first and second bandgap for Mode P with the variation of $\delta$, while Fig. 4 (c) shows the same process for the second to fourth bandgaps for Mode S. The scattered and solid lines represent the numerical (FE) and theoretical results of the edge frequencies of the bandgaps, respectively. The regions sandwiched by the curves indicate bandgaps, with colors indicating the bandgap sign $\varsigma$. Additionally, the colors of the lines represent the symmetric (green) or antisymmetric (grey) modes, and typical eigenmodes are presented in Fig. 4 (b) and (d) for longitudinal and transverse wave, respectively. The colors on the eigenmodes reflect the magnitude of displacement.

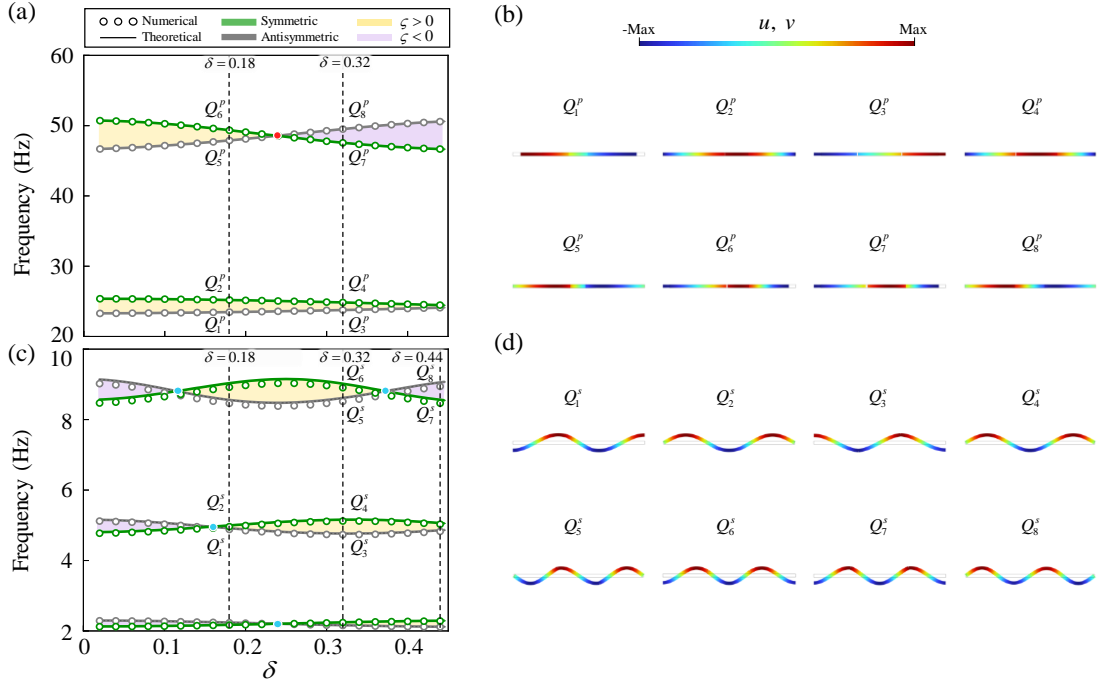

Fig. 4. Topological phase diagram of the soft Rayleigh beam with imperfect interfaces while varying $\delta$. (a) The evolution process of the first and second bandgap of Mode P, and (c) the same process of the second to fourth bandgaps of Mode S, with typical eigenmodes $Q_1^p \sim Q_8^p$, $Q_1^s \sim Q_8^s$ denoted in (b) and (d), respectively. The degeneracy position of the bandgap in (a) and (c) is marked in the red and bule dot, respectively. The position of the typical eigenmodes on the band structure in denoted in Fig. 2. The scattered and solid lines denote the numerical and theoretical results, respectively. The colors of the lines indicate the symmetric (green) or antisymmetric (gray) mode type. The colors on the eigenmodes reveal the magnitude of displacement.

As $\delta$ increases, the first bandgap of the longitudinal mode becomes smaller, but does not close. In addition, the mode of the lower boundary remains antisymmetric while the upper boundary remains symmetric, as illustrated in Fig. 4 (a). The bandgap sign also remains $\varsigma > 0$. However, the range of the second bandgap initially becomes smaller as $\delta$ increases from 0.02 to 0.24, and closes at $\delta = 0.24$. Afterward, the second bandgap reopens with further increases in $\delta$. For $\delta < 0.24$, the lower (higher) band is the antisymmetric (symmetric) mode, the bandgap sign is $\varsigma > 0$. Then, for $\delta > 0.24$, the symmetry of band state and the bandgap sign get inverted. These phenomena also can be confirmed through Fig. 4 (b). For the first bandgap, the lower (upper) boundary denoted $Q_1^p$ ($Q_2^p$) at $\delta = 0.18$ remain the antisymmetric (symmetric) type at $Q_3^p$ ($Q_4^p$) when $\delta = 0.32$. However, the mode type of the lower boundary of the second bandgap comes from an antisymmetric type at $Q_5^p$ to a symmetric type at $Q_7^p$, meanwhile the upper boundary comes from a symmetric type at $Q_6^p$ to an antisymmetric type at $Q_8^p$. This confirms the occurrence of the topological phase transition process.

Fig. 4 (c) illustrates a successive band state inversion process of the transverse wave. With the increment of $\delta$, the fourth bandgap first degenerates at $\delta = 0.115$, corresponding to 8.8 Hz, followed by the degeneration of the third bandgap at $\delta = 0.155$ (4.9 Hz). Finally, the second bandgap degenerates at $\delta = 0.24$ (2.2 Hz). Afterwards, the fourth bandgap degenerates again at $\delta = 0.375$ (8.8 Hz). The degeneracy of the bandgap is accompanied by the inversion of the symmetry type and the band sign, which is similar to the longitudinal wave discussed above, and can be confirmed through the eigenmodes presented in Fig. 4 (d). We conclude that the degeneracy of bandgap at higher frequency is more sensitive to the variation of $\delta$, thus experiencing more degeneracy events and degenerating at smaller $\delta$ compared to the bandgap at lower frequency. Additionally, the bandgap at higher frequency also has a wider range in general, which will be explained in detail in section 3.3.

TPIMs emerge at the interface of a mixed phononic crystal waveguide consisting of two elements that possess overlapped bandgaps with different topological properties, specifically, inverse bandgap signs. Motivated by this fact, a 20-cell beam system composed of 10 unit cells with geometric parameter $\delta_1$ and the other ten cells with $\delta_2$ is designed, as depicted in Fig. 5 (a). The transmission ratio, which is defined as

$$\xi = 20\log_{10}\left(\frac{|\theta|_{out}}{|\theta|_{in}}\right), \tag{49}$$

is evaluated to find the occurrence of TPIMs. We set $\theta = u$ for the longitudinal wave, and $\theta = v$ for the transverse wave. Fig. 5 (b) and (d) present the numerically calculated transmission ratios, while Fig. 5 (c) and (e) depict the typical mode shape of TPIM in the passband (green dot) and the bandgap (orange dot) region for Mode P and S, respectively. The height of the structure in Fig. 5 (c) and (e) is enlarged ten times for ease of observation, compared to its actual size. The colors on the eigenmodes represent the absolute magnitude of displacement.

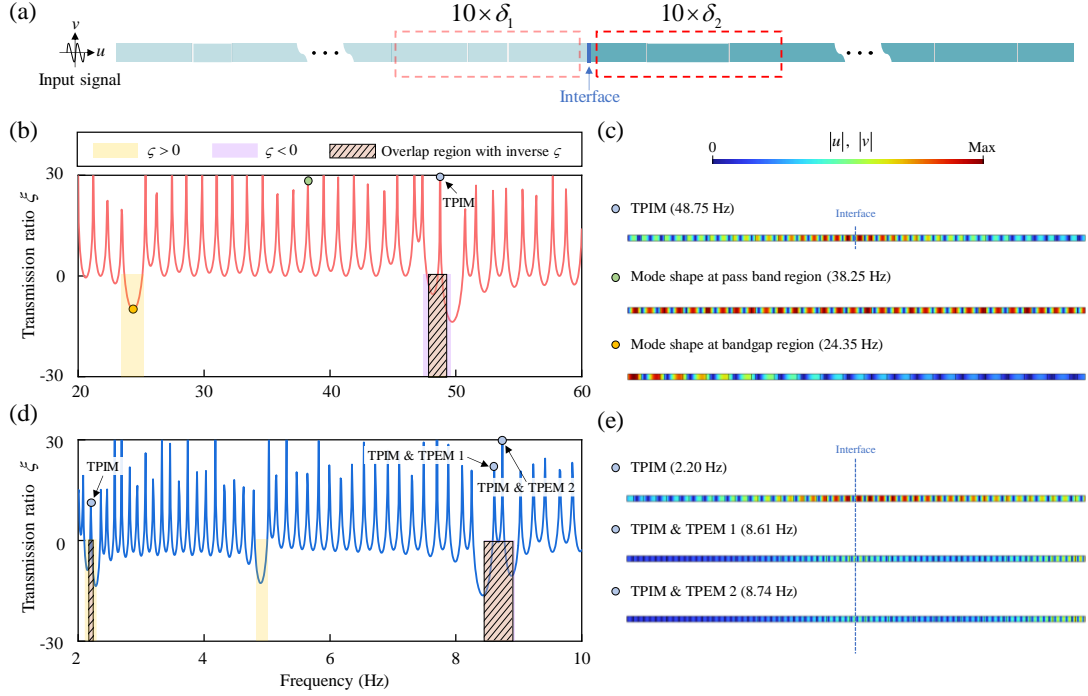

Fig. 5. Transmission ratio of the supercell containing 20 unit cells with different geometric parameters $\delta_1$ and $\delta_2$, as the schematic shown in (a). The numerical transmission ratio response for the supercell when $\delta_1 = 0.18$, $\delta_2 = 0.32$ is illustrated in (b) for longitudinal wave, while for transverse wave set $\delta_1 = 0.4$, $\delta_2 = 0.18$ in (d). The stripes in yellow and purple indicate the bandgap signs with $\varsigma > 0$ and $\varsigma < 0$, respectively, and the overlapped area with inverse $\varsigma$ is shown with black slant lines. The mode shape of TPIM or TPEM (blue dot), in the passband (green dot) and in the bandgap (orange dot) region is shown in (c) and (e) for the P and S mode, respectively. For ease of observation, the height of the structure in (c) and (e) is ten times the actual size. The colors on the eigenmodes denote the absolute magnitude of displacement.

We set $\delta_1 = 0.18$, $\delta_2 = 0.32$ to study the transmission effect of the longitudinal wave. The second bandgap exhibits an inverse bandgap sign, i.e., $\varsigma > 0$ at state $\delta_1 = 0.18$ with a frequency range of 47.89-49.34 Hz, while it becomes $\varsigma < 0$ at state $\delta_2 = 0.32$ with the frequency range of 47.55-49.50 Hz. The frequency range is got from the phase diagram Fig. 4. These two states have an overlapped bandgap region between 47.89-49.34 Hz, where the bandgap sign is opposite, thus a transmission peak representing the TPIM occurs at 48.75 Hz. Additionally, the transmission peak mode shape is extracted numerically and shown in Fig. 5 (c), where the displacement is localized near the interface between the two states of unit cells. In comparison, at 38.25 Hz, the displacement distribution is random in the passband for the entire structure, but in the bandgap region, it decays rapidly from the end at 24.35Hz.

In Fig. 5 (d) we set $\delta_1 = 0.44$, $\delta_2 = 0.18$ and investigate the transmission effect of the transverse wave. As shown in Fig. 4(c), there are two regions of bandgap sign inversion within the frequency range of 2.166-2.233 Hz and 8.166-8.919 Hz. Consequently, the TPIM emerge at 2.2 Hz. Coincidentally, at the range 8.166-8.919 Hz this system exhibits dual topologically protected modes, i.e., the interface mode and the edge mode. It is worth mentioning that the topologically protected edge mode (TPEM) is a special kind of TPIM, as the displacement is localized at the edge of the TI, which is also the "interface" between the TI and the vacuum. Two peaks in the transmission diagram occur at 8.61 and 8.74 Hz and their corresponding mode shapes are presented in Fig. 5 (e), where the effect of displacement localization at the center and edges of the beam system can be easily observed.

Contrary to the aforementioned discussion, the sign of the first and third bandgap in Fig. 5 (b) and (d) remains unchanged for the selected $\delta_1$ and $\delta_2$, respectively, resulting in a plain bandgap region with no transmission peak. This further underscores the significance of the bandgap sign. Consequently, by varying the geometric parameters of a Rayleigh beam system with imperfect interfaces, topological phase inversion can be induced, leading to the emergence of TPIMs for both longitudinal and transverse waves. The system can also support multiple TPIMs.

## *3.4. Effect of the imperfect interface and finite deformation*

To understand the influence of the imperfect interface and finite deformation on the band structure, based on the simplified numerical model, a variable separation type of research is carried out in this subsection. The topological phase diagrams for turning the stiffness of the imperfect interface and applying different finite deformations are presented in Fig. 6 and Fig. 7, respectively.

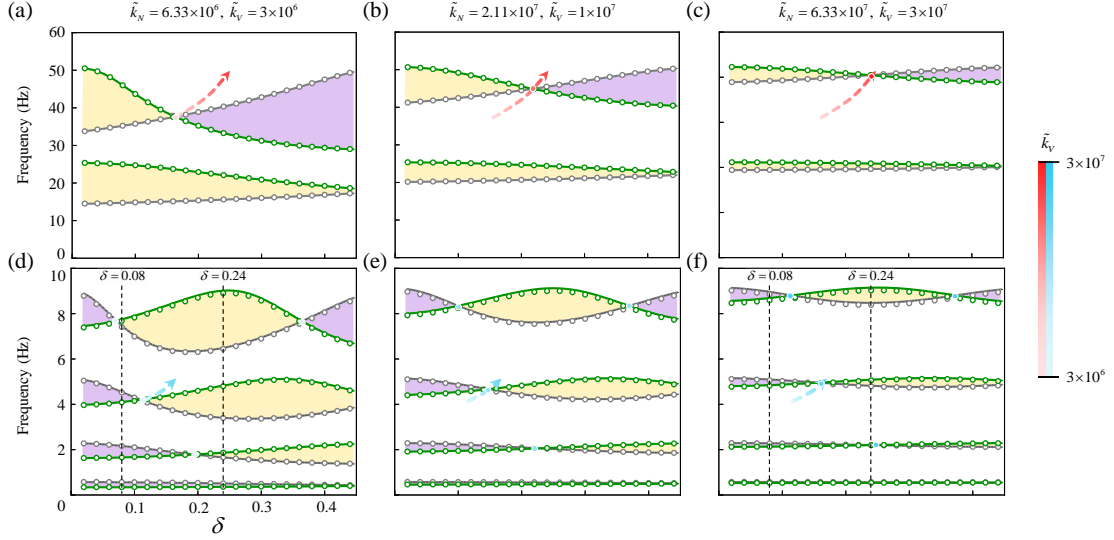

Fig. 6. Topological phase diagram of the soft Rayleigh beam with different stiffness of the imperfect interface (a)-(c) for Mode P and (d)-(f) for Mode S. Red and blue dashed arrows represent the locus of the topological transition point while altering $\tilde{k}_r$ $(r = M, N, V)$ for Mode P and S, respectively.

The topological phase diagrams under different stiffness of the imperfect interface are drawn Fig. 6 (a)-(c) and (d)-(f) for Mode P and S, respectively. The parameter $\lambda_1$ is kept in constant at 1 to isolate the effects of the imperfect interface. From Fig. 6, increasing the stiffness of the imperfect interface $\tilde{k}_N$ and $\tilde{k}_V$ leads to a decrease in bandgap width and a shift of bandgap position to higher frequencies. Additionally, the bandgap width tends to increase with frequency range, i.e., wider bandgaps at higher frequencies than at lower frequencies. These trends are consistent for both Mode P and S. Moreover, the frequencies of the topological phase transition points shift to higher frequencies, and the corresponding geometric parameter $\delta$ increases, especially for Mode P.

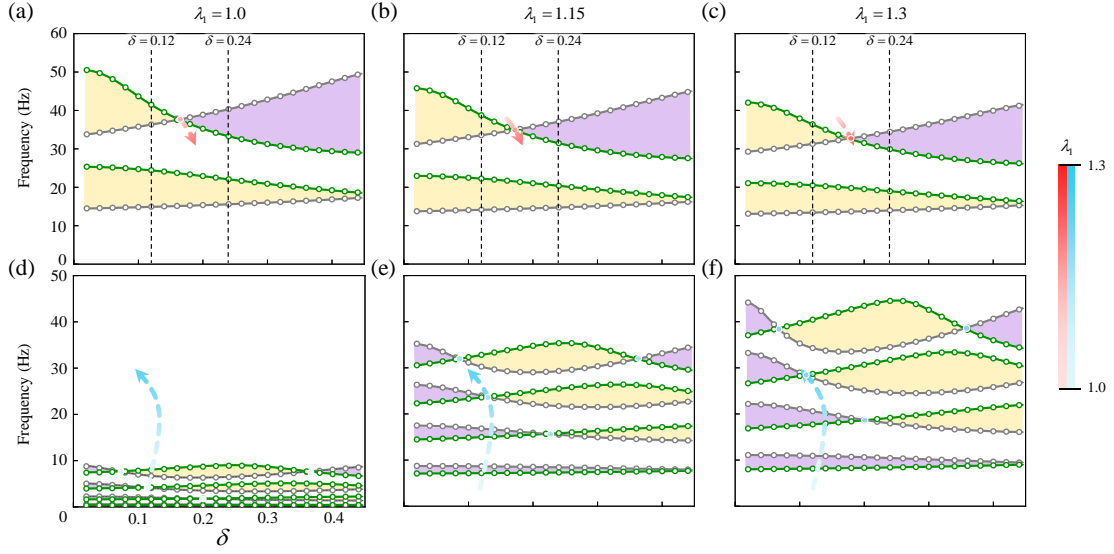

Fig. 7. Topological phase diagram of the soft Rayleigh beam with different finite deformations (a)-(c) for Mode P and (d)-(f) for Mode S. Red and blue dashed arrows represent the locus of the topological transition point while altering $\lambda_1$ for Mode P and S, respectively.

To study the effect of finite deformation on the beam system's band structure, the topological phase diagrams are drawn Fig. 7 (a)-(c) and (d)-(f) for Mode P and S, respectively, under uniaxial stretch $\lambda_1 = 1.0 \sim 1.3$. The parameter $\tilde{k}_N$ and $\tilde{k}_V$ are kept in constant at $6.33 \times 10^6 \ \mathrm{N/m^3}$ and $3 \times 10^6 \ \mathrm{N/m^3}$, respectively to isolate the effects of the finite deformation. Increasing the pre-deformation $\lambda_1$ leads to a shift of the bandgap regions of Mode P towards lower frequencies and a decrease in their widths. Moreover, the frequencies of the topological phase transition positions also shift to lower frequencies, and the corresponding geometric parameter $\delta$ increases. In contrast, the impact of uniaxial stretch $\lambda_1$ on Mode S is more pronounced, as the regions of all bandgaps of move towards higher frequencies and their widths increases, particularly from 1.0 to 1.15. Additionally, the frequencies of the topological phase transition positions shift to higher frequencies, albeit with a slight decrease in the corresponding geometric parameter $\delta$. These results still hold true when considering the effects of uniaxial stretch on the stiffness of imperfect interfaces, since $\tilde{k}_N$ and $\tilde{k}_V$ becomes almost unchanged when $\lambda_1 > 1.02$, as descripted in Supplementary Material.

The evolution of bandgaps and topological phase transition points while tuning $\tilde{k}_N$, $\tilde{k}_V$ and $\lambda_1$ can be explained through Eq. (40), which concludes that $\omega \propto 1/\eta_m$. After a uniaxial stretch along the $x$- direction, the eigenfrequency of Mode P decreases mainly affected by the increment of $1/\eta_p$, and the eigenfrequency of Mode S increases mainly by the increment of $1/\eta_s$. This result is consistent with prior research on a soft lattice (Zhao et al., 2022), and additional details can be found in the Supplementary Material. In brief, the uniaxial stretch primarily affects the geometrical and material parameters of the beam system, with distinct effects on Mode P and S.

Moreover, the expression $R_m(\omega,k)$ in Eq. (40) represents the dispersion effect, i.e., through solving $R_m(\omega,k)=0$ leads to the explicit line relation of $\omega$ and $k$, $\omega = c_m\left(\pm k + \dfrac{n}{l}\pi\right)$, $n \in \mathbb{Z}$, which is also the solution for the soft beam system without imperfect interfaces. On opposite, $T_m(\omega)$ represent the resonance effect, i.e., through solving $T_m(\omega)=0$ can get $\omega = \dfrac{n\pi c_m}{2l_t}$ or $\dfrac{n\pi c_m}{2(l-l_t)}$, $n \in \mathbb{Z}$, which are the eigenfrequencies of the beam with length $2l_t$ and $2(l-l_t)$, respectively, which is irrelevant with $k$ thus exhibits the flat bands on the band structure. The dimensionless coefficient $\kappa$ serves as the weighting factor between the dispersion effect and the resonance effect. Based on the formulation on $\kappa$ in Eq. (41), for any given wave velocity $c_m$ and density $\rho_0$, which represents any fixed deformed state, the weight coefficient $\kappa$ exhibits a positive correlation with the stiffness $\tilde{k}_m$, and has negative correlation with the angular frequency $\omega$.

An increase in $\kappa$ makes the beam system more susceptible to the dispersion effect, resulting in straighter band branches and narrower bandgaps at large $\tilde{k}_m$ or at lower frequency. Conversely, the increment of $\kappa$ makes the beam system more susceptible to the resonance effect, resulting in flattened band branches and larger bandgaps at small $\tilde{k}_m$ or at higher frequency. This explains the influence of $\tilde{k}_m$ on the band structure and the bandgap width distribution with frequency $\omega$. In conclusion, the parameter $\tilde{k}_m$ mainly affects the weights of the terms in Eq. (40), resulting in an identical effect on both Mode P and S. Such conclusions are deduced from the algebraic point of view, some additional explanations of these conclusions from the perspective of geometry are presented in the Supplementary Material.

## 3.5. Tunable TPIMs

The frequency of TPIMs can be adjusted using two primary methods: by applying uniaxial stretch or by adjusting the stiffness of the imperfect interface. In this section, two types of supercells are designed under different conditions to demonstrate the effectiveness of these methods and to verify the tunability of TPIMs in the soft beam system with imperfect interfaces.

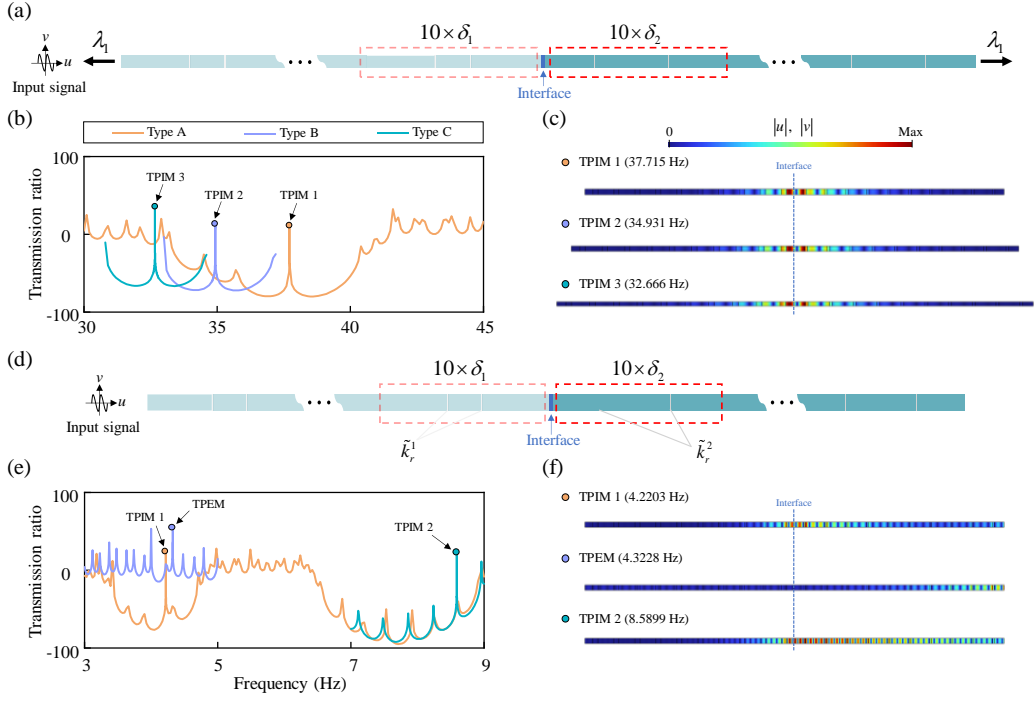

Fig. 8. Tunable TPIMs through various methods. (a) The schematic of the supercell modulated through finite deformation. (b) The corresponding transmission curve of Mode P. (c) The labeled mode shapes of TPIMs. (d) The schematic of the supercell modulated through different combinations of the stiffness of the imperfect interfaces $\tilde{k}_r^1$ and $\tilde{k}_r^2$ $(r = M, N, V)$. (e) The corresponding transmission curve of Mode S. (f) The labeled mode shapes of TPIMs and TPEM.

In the first case, set $\delta_1 = 0.24$, $\delta_2 = 0.12$, $\tilde{k}_N = 6.33 \times 10^6$ N/m$^3$, $\tilde{k}_V = 3 \times 10^6$, and apply different pre-deformation conditions with the schematic shown in Fig. 8 (a). The transmission ratio for Mode P is then analyzed and plotted in Fig. 8 (b). The solid lines in the figure are color-coded based on the magnitude of uniaxial stretch, where Type A, B, and C correspond to the values of $\lambda_1 = 1.0$, 1.15, 1.3, respectively. The results show that TPIMs exist in the frequency range of 30-45 Hz when the structure is stretched, and their mode shapes are displayed in Fig. 8 (c). The eigenfrequencies of these mode shapes are calculated and labeled, and are found to be in good agreement with the topological phase transition point shown in Fig. 7 (a)-(c). Additionally, the frequencies of the TPIMs decrease with increasing uniaxial stretch.

In the second case, set $\delta_1 = 0.24$, $\delta_2 = 0.08$, $\lambda_1 = 1$, and the transmission effect of the transverse wave is studied under different combinations of the stiffness of imperfect interface, as the schematic shown in Fig. 8 (d) and the results presented in Fig. 8 (e). Different from Fig. 8 (b), the colors of solid lines in Fig. 8 (e) represent different types of combinations. Type A: $\tilde{k}_N^1 = \tilde{k}_N^2 = 6.33 \times 10^6 \text{ N/m}^3$, $\tilde{k}_V^1 = \tilde{k}_V^2 = 3 \times 10^6 \text{ N/m}^3$. Type B: $\tilde{k}_N^1 = 6.33 \times 10^7 \text{ N/m}^3$, $\tilde{k}_V^1 = 3 \times 10^7 \text{ N/m}^3$, $\tilde{k}_N^2 = 6.33 \times 10^6 \text{ N/m}^3$, $\tilde{k}_V^2 = 3 \times 10^6 \text{ N/m}^3$. Type C: $\tilde{k}_N^1 = 6.33 \times 10^6 \text{ N/m}^3$, $\tilde{k}_V^1 = 3 \times 10^6 \text{ N/m}^3$, $\tilde{k}_N^2 = 6.33 \times 10^7 \text{ N/m}^3$, $\tilde{k}_V^2 = 3 \times 10^7 \text{ N/m}^3$. The comparison of Type A and B displays the switch from the interface mode to the edge mode, as shown in Fig. 8 (f), i.e., the TPIM at frequency 4.2203 Hz vanishes while increasing $\tilde{k}_N^1$ and $\tilde{k}_V^1$, and the TPEM appears at 4.3228 Hz, where the displacements are localized at the right end of the structure. Furthermore, there are no TPIMs in the frequency range 6-9 Hz for Type A because the bandgap sign is both positive both for $\delta_1$ and $\delta_2$, as shown in Fig. 6 (d). However, the bandgap sign becomes negative for $\delta_2$ while increasing $\tilde{k}_N^2$ and $\tilde{k}_V^2$, and an overlapped region for the inverse bandgap sign $\varsigma$ emerges, leading to the appearance of a TPIM at 8.5899 Hz. Consequently, the broad tunability of TPIMs in the soft beam system with imperfect interface, which includes altering the frequencies of the interface modes, switching from the interface mode to the edge mode, and opening up new TPIMs in the bandgap region through finite deformations and tuning the stiffness of the imperfect interface.

## 4. Conclusions

In this paper, the effects of uniaxial stretch and the imperfect interfaces on topological properties of a Rayleigh beam system with imperfect interfaces are studied. Theoretical models for calculating the stiffness of the soft imperfect interface and the dispersion relation of such beam system are established, while two types of numerical models employing FE method are adopted to elucidate the correction of the simplification of imperfect interfaces. The topological properties of the band structures are studied through numerical calculations of the Zak phase and bandgap sign. Specifically, the topological phase transition process is examined by adjusting the distance between imperfect interfaces. The transmission effect of a supercell comprising of two types of phononic crystal elements with overlapped bandgap region but different topological properties is calculated, verifying the existence of topologically protected interface modes. Additionally, the influence of finite deformation and the stiffness of imperfect interfaces on bandgaps and topological phase transition points is studied through topological phase diagrams. Finally, the wide tunability of TPIMs in this system is demonstrated using two numerical cases. Several valuable conclusions based on the numerical results and theoretical analyses are listed as follows:

➢ A soft topological insulator model that integrates the Rayleigh beam system with imperfect interfaces is established. This model facilitates the observation of the topological phase transition process by adjusting the distance between these imperfect interfaces. Moreover, the existence of topologically protected interface modes can be anticipated by analyzing the bandgap sign.

➢ A numerical simplification technique that reduces the soft thin layer to a linear spring-layer imperfect interface while studying the propagation of Bloch waves is proposed. The expression for the equivalent stiffness in the presence of finite deformation is provided through Eq. (25).

➢ There exist two methods to turn the band structure of this soft beam system: the application of finite deformation and manipulation of the imperfect interfaces' stiffness. As uniaxial stretch increases, bandgaps of Mode P and S display opposite movements, primarily dictated by the divergent alterations in $1/\eta_p$ and $1/\eta_s$, respectively. The stiffness of imperfect interfaces chiefly influences the dimensionless weight coefficient $\kappa$, transitioning the system from a dispersion type to a resonance type, producing analogous movements of the bandgaps for Modes P and S. Additionally, the bandgap width generally expands as the frequency range increases due to this effect.

➢ As inspired by the topological phase diagram, the topological phase transition point for Mode P can be adjusted within a narrow frequency range through the application of uniaxial stretch, whereas for Mode S, it can be adjusted within a broader frequency range. Conversely, by manipulating the stiffness of the imperfect interfaces, the topological phase transition point for Mode P can be adjusted within a large frequency range, while for Mode S, it can only be adjusted within a small frequency range.

➢ The soft beam system studied in this paper exhibits extensive tunability of TPIMs through finite deformation and adjusting the stiffness of the imperfect interface, including the ability to alter frequencies, switch from interface mode to edge mode, and even open up new TPIMs in a bandgap region.

This study emphasizes the significance of interface connection conditions, offering a novel approach and guideline for exploiting topological phase transition in composite continuous phononic crystal systems. These tunable interface states are highly desirable for scientists and engineers, and it is hoped that they can be extended to two-dimensional or even higher-order systems in the future.

## Declaration of Competing Interest

The authors declare that they have no known competing financial interests or personal relationships that could have appeared to influence the work reported in this paper.

## Acknowledgments

We acknowledge the support of the following: National Natural Science Foundation of China (Nos. 12102388, T2125009, 92048302), National Key R&D Program of China 2017 (YFA0701100), "Pioneer" R&D Program of Zhejiang (2023C03007) and Laoshan laboratory (LSKJ202205300), Key Research and Development Project of Zhejiang Province (2022C01022).

## Appendix A. Supplementary data

Supplementary data to this article can be found at the Supplementary Materials.